\documentclass[11pt]{article}
\usepackage{amsmath,amssymb,amsthm,amsxtra,overpic,bbm,bm,epsfig,ulem,color,multirow}
\usepackage{braket}

\begin{document}

\begin{center}
{\Large Leptogenesis implications of two-zero textures of\\
 Majorana neutrino mass matrix in type-I seesaw model }
\end{center}

\vspace{0.05cm}

\begin{center}
{\bf Xin-Tong Ma, Ding-Hui Xu, Zhen-hua Zhao\footnote{Corresponding author: zhaozhenhua@lnnu.edu.cn}, } \\
{ $^1$ Department of Physics, Liaoning Normal University, Dalian 116029, China \\
$^2$ Center for Theoretical and Experimental High Energy Physics, \\ Liaoning Normal University, Dalian 116029, China }
\end{center}

\vspace{0.2cm}

\begin{abstract}
The type-I seesaw model with two-zero textures of the Majorana mass matrix for the right-handed neutrinos $M^{}_{\rm R}$ provides a highly predictive framework for neutrino mass generation and  baryon asymmetry of the Universe via leptogenesis. In this work, we systematically investigate the compatibility of viable two-zero textures of $M^{}_{\rm R}$ with leptogenesis within the type-I seesaw scenario. We consider both a general diagonal Dirac neutrino mass matrix $M^{}_{\rm D}$ and two theoretically motivated special cases, namely the SO(10) GUT-inspired $M^{}_{\rm D} \sim \mathrm{diag}(m_u,m_c,m_t)$ and the flavor symmetry-induced $M^{}_{\rm D} \propto I$. Our study shows that two-zero textures of $M^{}_{\rm R}$ yield strong correlations between neutrino parameters and leptogenesis, offering distinctive phenomenological implications for neutrino flavor physics and baryon asymmetry of the Universe.
\end{abstract}

\newpage

\section{Introduction}

As we know, the phenomenon of neutrino oscillations indicates that neutrinos are massive and their flavor eigenstates $\nu^{}_\alpha$ (for $\alpha =e, \mu, \tau$) are certain superpositions of the mass eigenstates $\nu^{}_i$ (for $i =1, 2, 3$) with definite masses $m^{}_i$: $\nu^{}_\alpha = \sum^{}_i U^{}_{\alpha i} \nu^{}_i$ with $U^{}_{\alpha i}$ being the $\alpha i$-th element of the $3 \times 3$  neutrino mixing matrix $U$ \cite{xing}. In the case that neutrinos are Majorana particles, $U$ can be expressed in terms of three mixing angles $\theta^{}_{ij}$ (for $ij=12, 13, 23$), one Dirac CP phase $\delta$ and two Majorana CP phases $\rho$ and $\sigma$ in a way as follows:
\begin{eqnarray}
U  = \left( \begin{matrix}
c^{}_{12} c^{}_{13} & s^{}_{12} c^{}_{13} & s^{}_{13} e^{-{\rm i} \delta} \cr
-s^{}_{12} c^{}_{23} - c^{}_{12} s^{}_{23} s^{}_{13} e^{{\rm i} \delta}
& c^{}_{12} c^{}_{23} - s^{}_{12} s^{}_{23} s^{}_{13} e^{{\rm i} \delta}  & s^{}_{23} c^{}_{13} \cr
s^{}_{12} s^{}_{23} - c^{}_{12} c^{}_{23} s^{}_{13} e^{{\rm i} \delta}
& -c^{}_{12} s^{}_{23} - s^{}_{12} c^{}_{23} s^{}_{13} e^{{\rm i} \delta} & c^{}_{23}c^{}_{13}
\end{matrix} \right) \left( \begin{matrix}
e^{{\rm i}\rho} &  & \cr
& e^{{\rm i}\sigma}  & \cr
&  & 1
\end{matrix} \right) \;,
\label{1.1}
\end{eqnarray}
where the abbreviations $c^{}_{ij} = \cos \theta^{}_{ij}$ and $s^{}_{ij} = \sin \theta^{}_{ij}$ have been employed.

Among the nine observable parameters in the neutrino sector (i.e., three neutrino masses, three mixing angles and three CP phases), only 2 independent neutrino mass-squared differences $\Delta m^2_{ij} \equiv m^2_i - m^2_j$ and three mixing angles have been measured to a good degree of accuracy. And there is a preliminary result for $\delta$ but with large uncertainties, so we will treat it as a free parameter. The global analysis of neutrino oscillation data is available in Refs.~\cite{global1}-\cite{global3}, and we will use the results in Ref.~\cite{global1} (shown in Table~1 here) as reference values in the following numerical calculations.
However, the values of the lightest neutrino mass and Majorana CP phases are completely unknown. Furthermore, there is an ambiguity for the neutrino mass ordering---two possibilities are allowed: the normal ordering (NO) case with $m^{}_1 < m^{}_2 < m^{}_3$, or the inverted ordering (IO) case with $m^{}_3 < m^{}_1 < m^{}_2$.

In the lack of a convincing flavor theory, many approaches (e.g., flavor symmetries or texture zeros) have been tried to study the flavor problems of massive neutrinos. In the texture-zero approach, the Majorana neutrino mass matrix $M^{}_\nu$ may have a few texture zeros. And it has been pointed out that such texture zeros may naturally arise from underlying Abelian flavor symmetries \cite{zero}. These texture zeros are phenomenologically useful in the sense that they can help us to establish some simple and testable relations between neutrino mixing parameters and neutrino mass ratios---if such relations turn out to be favored by the experimental data, they might have a fundamental reason and should originate from the underlying flavor theory. Hence a phenomenological study of possible texture zeros of $M^{}_\nu$ makes sense. Given that $M^{}_\nu$ is symmetric, it has six independent complex entries. It is easy to show that the textures of $M^{}_\nu$ with more than two independent zeros are definitely incompatible with current neutrino oscillation data \cite{3zero}. Hence most of the relevant studies have focused on the two-zero textures of $M^{}_\nu$ (for an incomplete list of related references, see Refs.~\cite{2zero1}-\cite{2zero12}). Among the fifteen (i.e., $C^2_6$) possible two-zero texture classes of $M^{}_\nu$, only six are still consistent with present neutrino oscillation data, which are listed in the left panel of Table~2 \cite{2zero5}-\cite{2zero12}.

\begin{table}\centering
  \begin{footnotesize}
    \begin{tabular}{c|cc|cc}
     \hline\hline
      & \multicolumn{2}{c|}{Normal Ordering}
      & \multicolumn{2}{c}{Inverted Ordering }
      \\
      \cline{2-5}
      & bfp $\pm 1\sigma$ & $3\sigma$ range
      & bfp $\pm 1\sigma$ & $3\sigma$ range
      \\
      \cline{1-5}
      \rule{0pt}{4mm}\ignorespaces
       $\sin^2\theta^{}_{12}$
      & $0.303_{-0.011}^{+0.012}$ & $0.270 \to 0.341$
      & $0.303_{-0.011}^{+0.012}$ & $0.270 \to 0.341$
      \\[1mm]
       $\sin^2\theta^{}_{23}$
      & $0.572_{-0.023}^{+0.018}$ & $0.406 \to 0.620$
      & $0.578_{-0.021}^{+0.016}$ & $0.412 \to 0.623$
      \\[1mm]
       $\sin^2\theta^{}_{13}$
      & $0.02203_{-0.00059}^{+0.00056}$ & $0.02029 \to 0.02391$
      & $0.02219_{-0.00057}^{+0.00060}$ & $0.02047 \to 0.02396$
      \\[1mm]
       $\delta/\pi$
      & $1.09_{-0.14}^{+0.23}$ & $0.96 \to 1.33$
      & $1.59_{-0.18}^{+0.15}$ & $1.41\to 1.74$
      \\[3mm]
       $\Delta m^2_{21}/(10^{-5}~{\rm eV}^2)$
      & $7.41_{-0.20}^{+0.21}$ & $6.82 \to 8.03$
      & $7.41_{-0.20}^{+0.21}$ & $6.82 \to 8.03$
      \\[3mm]
       $\Delta m^2_{3\ell}/(10^{-3}~{\rm eV}^2)$
      & $2.511_{-0.027}^{+0.028}$ & $2.428 \to 2.597$
      & $-2.498_{-0.025}^{+0.032}$ & $-2.581 \to -2.408$
      \\[2mm]
      \hline\hline
    \end{tabular}
  \end{footnotesize}
  \caption{The best-fit values, 1$\sigma$ errors and 3$\sigma$ ranges of six neutrino
oscillation parameters extracted from a global analysis of the existing
neutrino oscillation data \cite{global1}, where $ m^2_{3\ell}= m^2_{31}>0$ for NO and $ m^2_{3\ell}= m^2_{32}>0$ for IO. }
\end{table}

On the other hand, one of the most popular and natural ways of generating the neutrino masses is the type-I seesaw model in which heavy right-handed neutrinos $N^{}_I$ ($I=1, 2, 3$) are introduced into the Standard Model (SM) \cite{seesaw1}-\cite{seesaw5}. Like other fermions, the right- and left-handed neutrinos can constitute the Yukawa coupling operators together with the Higgs doublet $H$: $(Y^{}_{\nu})^{}_{\alpha I} \overline {L^{}_\alpha} H N^{}_I $ with $L^{}_\alpha$ being the left-handed lepton doublets. These operators will generate the Dirac neutrino masses $(M^{}_{\rm D})^{}_{\alpha I}= (Y^{}_{\nu})^{}_{\alpha I} v$ after the neutral component of $H$ acquires the nonzero vacuum expectation value (VEV) $v = 174$ GeV. Furthermore, $N^{}_I$ themselves can also have the Majorana mass terms $\overline{N^c_I} (M^{}_{\rm R})^{}_{IJ} N^{}_J$.
Then, under the seesaw condition $M^{}_{\rm R} \gg M^{}_{\rm D}$, integrating the right-handed neutrinos out will yield the aforementioned Majorana mass matrix $M^{}_\nu$ for the three light neutrinos:
\begin{eqnarray}
M^{}_{\nu} \simeq - M^{}_{\rm D} M^{-1}_{\rm R} M^{T}_{\rm D} \;.
\label{1.2}
\end{eqnarray}

\begin{table}\centering
    \begin{tabular}{|cc|cc|}
     \hline
   Class  & Texture Zeros &
   Class  & Zero Cofactors
      \\
    \hline
   $A^{}_1$ & $M^{}_{ee}$ \& $M^{}_{e\mu}$
      & $A^{}_1$ & $C^{}_{\mu\tau}$  \& $C^{}_{\tau\tau}$
      \\[1mm]
   $A^{}_2$ & $M^{}_{ee}$  \& $M^{}_{e\tau}$
      & $A^{}_2$ & $C^{}_{\mu\mu}$  \& $C^{}_{\mu\tau}$
      \\[1mm]
   $B^{}_1$ & $M^{}_{e\tau}$  \& $M^{}_{\mu\mu}$
      & $B^{}_3$ & $C^{}_{e\tau}$  \& $C^{}_{\tau\tau}$
      \\[1mm]
   $B^{}_2$ & $M^{}_{e\tau}$  \& $M^{}_{\tau\tau}$
      & $B^{}_4$ & $C^{}_{e\mu}$  \& $C^{}_{\mu\mu}$
      \\[1mm]
   $B^{}_3$ & $M^{}_{e\tau}$  \& $M^{}_{\mu \mu}$
      & $B^{}_6$ & $C^{}_{e\tau}$  \& $C^{}_{\mu\mu}$
      \\[1mm]
   $B^{}_4$ & $M^{}_{e\tau}$  \& $M^{}_{\tau \tau}$
      & - & -
       \\[1mm]
      \hline
    \end{tabular}
  \caption{ Experimentally allowed classes of two-zero textures (left panel) and two-zero cofactors (right panel) of $M^{}_\nu$ \cite{2zero7}. Here $M^{}_{\alpha\beta}$ and $C^{}_{\alpha\beta}$ respectively denote the texture zero and zero cofactors for the $\alpha \beta$-th element of $M^{}_\nu$. }
\end{table}

Remarkably, as an extra bonus, the seesaw model also provides an attractive explanation (known as the leptogenesis mechanism \cite{leptogenesis}-\cite{Lreview4}) for the baryon-antibaryon asymmetry of the Universe \cite{planck}
\begin{eqnarray}
Y^{}_{\rm B} \equiv \frac{n^{}_{\rm B}-n^{}_{\rm \bar B}}{s} \simeq (8.69 \pm 0.04) \times 10^{-11}  \;,
\label{1.3}
\end{eqnarray}
where $n^{}_{\rm B}$ ($n^{}_{\rm \bar B}$) denotes the baryon (antibaryon) number density and $s$ is the entropy density. The leptogenesis mechanism works elegantly in a way as follows: a lepton-antilepton asymmetry $Y^{}_{\rm L} \equiv (n^{}_{\rm L} - n^{}_{\rm \bar L})/s$ is first generated from the out-of-equilibrium and CP-violating decays of right-handed neutrinos and then partly converted into the baryon-antibaryon asymmetry via the sphaleron processes: $Y^{}_{\rm B} \simeq cY^{}_{\rm L}$ with $c = -28/79$.

Based on the above facts, this paper aims to investigate the implications of two-zero textures of Majorana neutrino mass matrix for leptogenesis within the full seesaw framework. Specifically, we impose texture zeros on the Majorana mass matrix $M^{}_{\rm R}$ of right-handed neutrinos, in the basis where the Dirac neutrino mass matrix $M^{}_{\rm D}$ is diagonal. The motivation of this study can be summarized as follows: within the seesaw model, $M^{}_{\rm R}$ is more fundamental than $M^{}_\nu$, making it more physically reasonable to impose zero textures on $M^{}_{\rm R}$ rather than on $M^{}_\nu$ \cite{MR}. In the full seesaw framework, we can further investigate the implications of texture zeros in the Majorana neutrino mass matrix for leptogenesis, thus distinguishing the compatibility of different texture ansatze with the leptogenesis mechanism.

To be concrete, we first consider the scenario where $M^{}_{\rm D}$ is a general diagonal matrix. We then examine two well-motivated special cases: $M^{}_{\rm D} \sim {\rm diag}(m^{}_u, m^{}_c, m^{}_t)$ (where $m^{}_{u, c, t}$ are the up, charm, and top quark masses), motivated by SO(10) grand unified theory (GUT) constructions (see, e.g., Ref.~\cite{GUT}), and $M^{}_{\rm D} \propto I$ (where $I$ denotes the identity matrix), inspired by non-Abelian flavor symmetries (see, e.g., Ref.~\cite{FS}). The goal of investigating these two particular cases is to explore the leptogenesis implications of two-zero textures within the SO(10) GUT \cite{universal20} and flavor symmetry frameworks.

The remaining parts of this paper are organized as follows. In sections~2-4, we successively investigate the cases that $M^{}_{\rm D}$ is a general diagonal matrix, $M^{}_{\rm D} \sim {\rm diag}(m^{}_u, m^{}_c, m^{}_t)$, and $M^{}_{\rm D} \propto I$, respectively. Section~5 summarizes our main conclusions.

\section{$M^{}_{\rm D}$ as a general diagonal matrix}

We first consider the case that $M^{}_{\rm D}$ is a general diagonal matrix parameterized as
\begin{eqnarray}
M^{}_{\rm D} = k \left( \begin{matrix}
1 &  &  \cr
 & r^{}_2  &  \cr
 &  & r^{}_3
\end{matrix} \right) \;.
\label{2.1}
\end{eqnarray}
Here $k$ denotes the overall scale of $M^{}_{\rm D}$, while $r^{}_2$ and $r^{}_3$ are dimensionless parameters. For a diagonal $M^{}_{\rm D}$, texture zeros of $M^{}_{\rm R}$ manifest themselves as zero cofactors (or equivalently, zero minors) of $M^{}_\nu$ \cite{MR, 2minor, 2cofactor}. Since $M^{}_\nu$ can be reconstructed from the neutrino mixing matrix $U$ in Eq.~(\ref{1.1}) and three neutrino masses through the relation
\begin{eqnarray}
M^{}_{\nu} = U D^{}_\nu U^T \;,
\label{2.2}
\end{eqnarray}
with $D^{}_\nu = {\rm diag}(m^{}_1, m^{}_2, m^{}_3)$, a zero cofactor for the $\alpha \beta$-th element of $M^{}_\nu$ will yield the following constraint equations:
\begin{eqnarray}
&& C^{}_{ee} = 0 \hspace{0.5cm} \longrightarrow \hspace{0.5cm} M^{}_{\mu\mu} M^{}_{\tau\tau} - M^{}_{\mu\tau} M^{}_{\mu\tau} =0 \;, \nonumber\\
&& C^{}_{e\mu} = 0 \hspace{0.5cm} \longrightarrow \hspace{0.5cm} M^{}_{e\mu} M^{}_{\tau\tau} - M^{}_{e\tau} M^{}_{\mu\tau} =0 \;, \nonumber\\
&& C^{}_{e\tau} = 0 \hspace{0.5cm} \longrightarrow \hspace{0.5cm} M^{}_{e\mu} M^{}_{\mu\tau} - M^{}_{e\tau} M^{}_{\mu\mu} =0 \;, \nonumber\\
&& C^{}_{\mu\mu} = 0 \hspace{0.5cm} \longrightarrow \hspace{0.5cm} M^{}_{ee} M^{}_{\tau\tau} - M^{}_{e\tau} M^{}_{e\tau} =0 \;, \nonumber\\
&& C^{}_{\mu\tau} = 0 \hspace{0.5cm} \longrightarrow \hspace{0.5cm} M^{}_{ee} M^{}_{\mu\tau} - M^{}_{e\mu} M^{}_{e\tau} =0 \;, \nonumber\\
&& C^{}_{\tau\tau} = 0 \hspace{0.5cm} \longrightarrow \hspace{0.5cm} M^{}_{ee} M^{}_{\mu\mu} - M^{}_{e\mu} M^{}_{e\mu} =0 \;,
\label{2.3}
\end{eqnarray}
with the $\alpha \beta$-th element of $M^{}_\nu$ explicitly expressed as
\begin{eqnarray}
M^{}_{\alpha \beta} = m^{}_1 U^{}_{\alpha 1} U^{}_{\beta 1} + m^{}_2 U^{}_{\alpha 2} U^{}_{\beta 2} + m^{}_3 U^{}_{\alpha 3} U^{}_{\beta 3} \;.
\label{2.4}
\end{eqnarray}
Furthermore, two-zero cofactors of $M^{}_\nu$ will give rise to two constrain equations from Eq.~(\ref{2.3}) for the neutrino mixing matrix elements and neutrino masses. As the constraint equations in Eq.~(\ref{2.3}) are complex, two such equations effectively correspond to four real constraint equations. In this case, substituting the global fit results of the neutrino mass-squared differences and neutrino mixing angles, two-zero cofactors of $M^{}_\nu$ allow us to fix the four undetermined neutrino parameters: the lightest neutrino mass and the three CP phases.

Our numerical analysis reveals that, among the fifteen (i.e., $C^2_6$) possible two-zero cofactor classes of $M^{}_\nu$, only five remain compatible with the neutrino oscillation data within their $3\sigma$ ranges, which are listed in the right panel of Table~2.
Among these five classes, $A^{}_1$, $A^{}_2$ and $B^{}_6$ are viable only for the normal ordering case of neutrino masses, while $B^{}_3$ and  $B^{}_4$ are viable for both the normal and inverted ordering cases. Note that some of these classes ($A^{}_{1}$, $A^{}_2$, $B^{}_{3}$ and $B^{}_4$) are equivalent to the two-zero textures of $M^{}_\nu$. Specifically, in the full type-I seesaw model with a diagonal $M^{}_{\rm D}$, these five classes correspond to the following two-zero textures of $M^{}_{\rm R}$:
\begin{eqnarray}
&& A^{}_1 : \hspace{0.2cm} M^{}_{\rm R}  = \left( \begin{matrix}
\times & \times & \times \cr
\times & \times  & 0 \cr
\times & 0 & 0
\end{matrix} \right) \;; \hspace{1cm}
A^{}_2 : \hspace{0.2cm} M^{}_{\rm R}  = \left( \begin{matrix}
\times & \times & \times \cr
\times & 0  & 0 \cr
\times & 0 & \times
\end{matrix} \right) \;; \nonumber \\
&& B^{}_3 : \hspace{0.2cm} M^{}_{\rm R}  = \left( \begin{matrix}
\times & \times & 0 \cr
\times & \times  & \times \cr
0 & \times & 0
\end{matrix} \right) \;; \hspace{1cm}
B^{}_4 : \hspace{0.2cm} M^{}_{\rm R}  = \left( \begin{matrix}
\times & 0 & \times \cr
0 & 0  & \times \cr
\times & \times & \times
\end{matrix} \right) \;; \nonumber \\
&& B^{}_6 : \hspace{0.2cm} M^{}_{\rm R}  = \left( \begin{matrix}
\times & \times & 0 \cr
\times &  0  & \times \cr
0 & \times & \times
\end{matrix} \right) \;.
\label{2.5}
\end{eqnarray}

For the above five classes, Figure~1 has shown their predictions for the lightest neutrino mass ($m^{}_1$/$m^{}_3$ in the NO/IO case) and $\delta$. We see that the results for $A^{}_1$ and $A^{}_2$ are similar: $m^{}_1$ is predicted to be around 0.005 eV, while $\delta$ is allowed to span a wide range from $\pi$ to $2\pi$. And the results for $B^{}_3$, $B^{}_4$ and $B^{}_6$ share similarities: the lightest neutrino mass is predicted to be close to 0.1 eV, while $\delta$ is close to $1.5 \pi$. Further, Figure~2 has shown the corresponding predictions for the Majorana CP phases $\rho$ and $\sigma$ for these five classes. For $A^{}_1$ and $A^{}_2$, their results are again similar: $\rho$ and $\sigma$ satisfy the relations $\rho \sim \sigma + \pi/2$ or $\sigma \sim \rho + \pi/2$. And the results for $B^{}_3$, $B^{}_4$ and $B^{}_6$ are also similar: $\rho$ and $\sigma$ are both predicted to be around $\pi/2$.

\begin{figure*}
\centering
\includegraphics[width=6.5in]{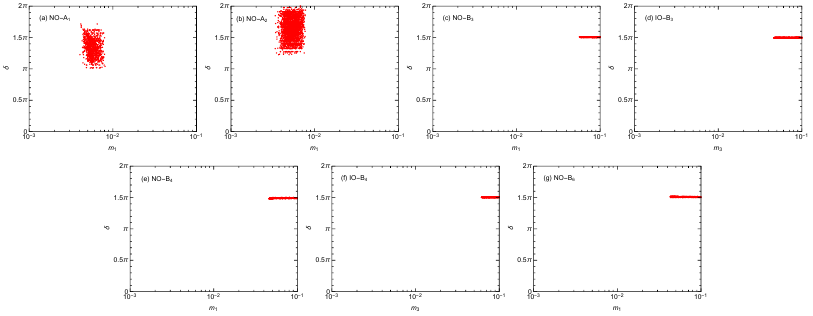}
\caption{ For the five experimentally allowed two-zero cofactor classes of $M^{}_\nu$ studied in section~2, their predictions for the lightest neutrino mass ($m^{}_1$/$m^{}_3$ in the NO/IO case) and $\delta$. }
\end{figure*}

\begin{figure*}
\centering
\includegraphics[width=6.5in]{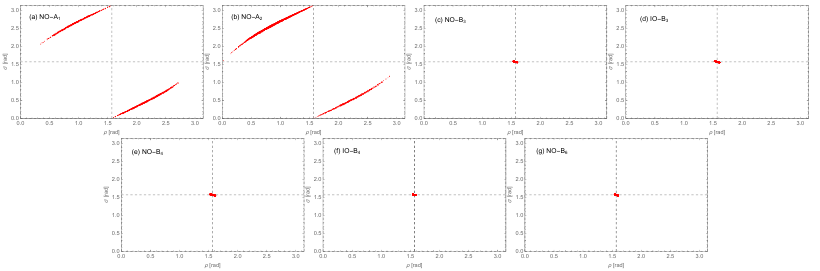}
\caption{ For the five experimentally allowed two-zero cofactor classes of $M^{}_\nu$ studied in section~2, their predictions for the Majorana CP phases $\rho$ and $\sigma$. }
\end{figure*}

Before proceeding to the leptogenesis calculations, we first examine the predictions for the lightest right-handed neutrino mass in the above five classes. By inverting Eq.~(\ref{1.2}), one can derive $M_{\rm R}$ from $M_\nu$ and $M_{\rm D}$ as
\begin{eqnarray}
M_{\rm R} \simeq - M_{\rm D}^T M_\nu^{-1} M_{\rm D} \;.
\label{2.6}
\end{eqnarray}
Diagonalizing the resulting $M^{}_{\rm R}$ via a unitary transformation $V^{}_{\rm R}$ yields the masses of three right-handed neutrinos (ordered by increasing mass). Since $M^{}_\nu$ is fully determined by the two-zero cofactor conditions, $M_{\rm R}$ depends only on the free parameters in $M^{}_{\rm D}$ (i.e., $k$, $r^{}_2$ and $r^{}_3$ in Eq.~(\ref{2.1})). In Figure 3, we display the allowed values of the lightest right-handed neutrino mass $M^{}_1$ as functions of $r^{}_3$ in the naturally expected range $0.1 -10$ for several benchmark values of $r^{}_2$ (also in the range $0.1 -10$), with $k = 100$ GeV adopted as a benchmark scale. This choice is motivated by the fact that $M^{}_{\rm D}$ arises as the product of the Higgs VEV and the neutrino Yukawa couplings. But it is worth noting that $M^{}_{\rm R}$ and the corresponding right-handed neutrino masses scale as $k^2$, so the results for other values of $k$ can be easily obtained from Figure~3 by an appropriate rescaling. The reason we focus on $M^{}_1$ is that the final baryon asymmetry generated via leptogenesis is dominated by the decay of the lightest right-handed neutrino and is positively correlated with $M^{}_1$. The results show that $M^{}_1$ generally increases with $r^{}_2$ and $r^{}_3$, a behavior that can be easily understood with the help of Eq.~(\ref{2.6}). Specifically, for the benchmark value of $k = 100$ GeV, $M^{}_1$ typically lies in the range $10^9 - 10^{11}$ GeV for $A^{}_1$ and $A^{}_2$, $10^{10} - 10^{12}$ GeV for $B^{}_3$ (in the NO case) and $B^{}_4$, and $10^{12} - 10^{14}$ GeV for $B^{}_3$ (in the IO case) and $B^{}_6$.

\begin{figure*}
\centering
\includegraphics[width=6.5in]{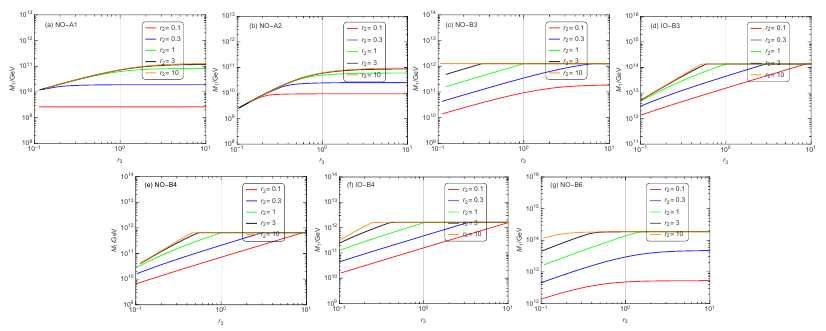}
\caption{ For the five experimentally allowed two-zero textures of $M^{}_{\rm R}$ studied in section~2, their predictions for the lightest right-handed neutrino mass $M^{}_1$ as functions of $r^{}_3$ for some benchmark values of $r^{}_2$. }
\end{figure*}

We now investigate the implications of the above five classes for leptogenesis. For the scenario with a non-diagonal $M^{}_{\rm R}$, it is more convenient to perform the leptogenesis calculations in the basis where $M^{}_{\rm R}$ is diagonalized by a unitary transformation $V^{}_{\rm R}$. In this basis, the Dirac neutrino mass matrix is rotated to $M^{\prime}_{\rm D} = M^{}_{\rm D} V^{}_{\rm R}$. The final baryon asymmetry generated from the decay of the lightest right-handed neutrino $N^{}_1$ can then be expressed as a product of several factors as
\begin{eqnarray}
Y^{}_{\rm B} = c r \varepsilon^{}_1 \kappa(K^{}_1)  \;.
\label{2.7}
\end{eqnarray}
Here $c \simeq -1/3$ is the conversion efficiency from the lepton asymmetry to the baryon asymmetry via the sphaleron processes, and $r \simeq 4 \times 10^{-3}$ measures the ratio of the equilibrium number density of $N^{}_1$ to the entropy density. $\varepsilon^{}_1$ denotes the total CP asymmetry (i.e., $\varepsilon^{}_1 = \sum^{}_\alpha \varepsilon^{}_{1 \alpha}$) associated with the decay of $N^{}_1$, with the flavor-specific CP asymmetry $\varepsilon^{}_{1 \alpha}$ explicitly given by
\begin{eqnarray}
&& \varepsilon^{}_{1 \alpha} = \frac{1}{8\pi (M^{\prime\dagger}_{\rm D}
M^{\prime}_{\rm D})^{}_{11} v^2} \sum^{}_{I \neq 1} \left\{ {\rm Im}\left[(M^{\prime *}_{\rm D})^{}_{\alpha 1} (M^{\prime}_{\rm D})^{}_{\alpha I}
(M^{\prime\dagger}_{\rm D} M^{\prime}_{\rm D})^{}_{1I}\right] {\cal F} \left( \frac{M^2_I}{M^2_1} \right) \right. \nonumber \\
&& \hspace{1.cm}
+ \left. {\rm Im}\left[(M^{\prime*}_{\rm D})^{}_{\alpha 1} (M^{\prime}_{\rm D})^{}_{\alpha I} (M^{\prime\dagger}_{\rm D} M^{\prime}_{\rm D})^*_{1I}\right] {\cal G}  \left( \frac{M^2_I}{M^2_1} \right) \right\} \; ,
\label{2.8}
\end{eqnarray}
where the loop functions are given by ${\cal F}(x) = \sqrt{x} \{(2-x)/(1-x)+ (1+x) \ln [x/(1+x)] \}$ and ${\cal G}(x) = 1/(1-x)$. Finally, $\kappa(K^{}_1)$ denotes the efficiency factor, corresponding to the survival probability of the lepton asymmetry produced in the decays of $N^{}_1$, which accounts for washout effects induced by the inverse decays of $N^{}_1$ and various lepton-number-violating scattering processes. Its value depends on the washout parameters
\begin{eqnarray}
K^{}_1 = \sum^{}_\alpha K^{}_{1\alpha} = \sum^{}_\alpha \frac{|(M^\prime_{\rm D})^{}_{\alpha 1}|^2}{M^{}_1 m^{}_*} \;,
\label{2.9}
\end{eqnarray}
with $m^{}_* \simeq  1.08 \times 10^{-3}$ eV. It is well approximately by
\begin{eqnarray}
\kappa(x)&=&\frac{2}{x z_{\rm B}(x)}\left[1-{\rm exp}\left(-\frac{1}{2} x\, z_{\rm B}(x)\right)\right]\,,\nonumber\\
z_{\rm B}(x)&\simeq& 2+4 \,x^{0.13} \,{\rm exp}\left(-\frac{2.5}{x}\right) \;.
\label{2.10}
\end{eqnarray}

It should be emphasized that Eq.~(\ref{2.7}) is valid only for leptogenesis occurring above $10^{12}$ GeV, where the charged-lepton Yukawa interactions are out of thermal equilibrium and lepton flavors are therefore indistinguishable. By contrast, when leptogenesis proceeds below $10^{12}$ GeV, the charged-lepton Yukawa interactions enter thermal equilibrium, and the individual lepton flavors become distinguishable and must be treated separately \cite{flavor1, flavor2}.
In the two-flavor regime realized for temperatures in the range $10^{9}-10^{12}$ GeV, the $\tau$ lepton flavor is distinguishable from the other two flavors. In this regime, the baryon asymmetry generated from $N^{}_1$ is given by
\begin{eqnarray}
Y^{}_{\rm B} =  c r \left[ (\varepsilon^{}_{1 e} + \varepsilon^{}_{1 \mu}) \kappa(K^{}_{1e} + K^{}_{1\mu}) + \varepsilon^{}_{1 \tau} \kappa(K^{}_{1\tau}) \right] \;.
\label{2.11}
\end{eqnarray}
In the three-flavor regime realized for temperatures below $10^{9}$ GeV, all the three lepton flavors are distinguishable from one another. In this regime, the baryon asymmetry generated from $N^{}_1$ is given by
\begin{eqnarray}
Y^{}_{\rm B} =  c r \left[ \varepsilon^{}_{1 e} \kappa(K^{}_{1e}) + \varepsilon^{}_{1 \mu} \kappa(K^{}_{1\mu}) + \varepsilon^{}_{1 \tau} \kappa(K^{}_{1\tau}) \right]  \;.
\label{2.12}
\end{eqnarray}

\begin{figure*}
\centering
\includegraphics[width=6.5in]{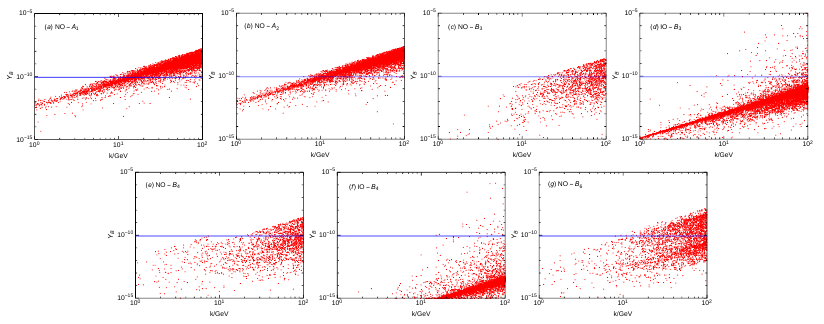}
\caption{ For the five experimentally allowed two-zero textures of $M^{}_{\rm R}$ studied in section~2, the allowed values of $Y^{}_{\rm B}$ as functions of $k$ for $r^{}_2$ and $r^{}_3$ in the range 0.1$-$10. The blue horizontal line stands for the observed value of $Y^{}_{\rm B}$. }
\end{figure*}

Based on the above formulas, for the above five classes, in Figure~4 we have shown the allowed values of $Y_{\rm B}$ as functions of $k$ for $r_2$ and $r_3$ in the range $0.1$--$10$. The results show that only for $k \gtrsim 10$ GeV can leptogenesis reproduce the observed baryon asymmetry. In Figure~5, we have further shown the values of $r_2$ versus $r_3$ that lead to successful leptogenesis for several benchmark values of $k$. The results show that viable parameter points yielding the observed baryon asymmetry exhibit distinct patterns across different two-zero texture classes: for $A^{}_1$ and $A^{}_2$, successful leptogenesis typically requires both $r^{}_2$ and $r^{}_3$ to be relatively large, at the order of 10. For $B^{}_3$ and $B^{}_4$ in the NO case, one of the parameters $r^{}_2$ or $r^{}_3$ lies close to 1 while the other is greater than 1. In contrast, for $B^{}_3$ and $B^{}_4$ in the IO case, one parameter is smaller than 1 and the other is larger than 1. As for $B^{}_6$, only the configuration with $r^{}_2 > 1$ and $r^{}_3 < 1$ can accommodate the observed baryon asymmetry.

\begin{figure*}
\centering
\includegraphics[width=6.5in]{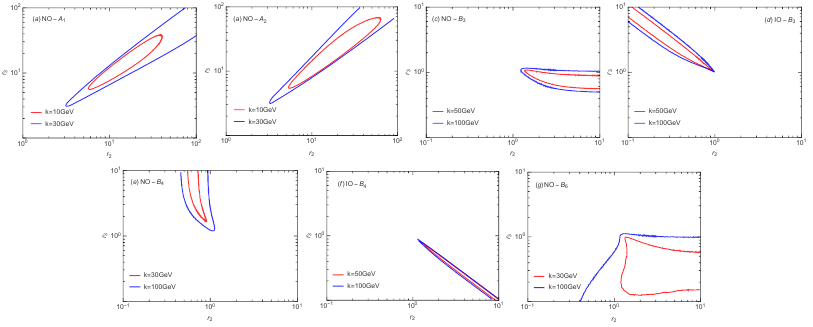}
\caption{ For the five experimentally allowed two-zero textures of $M^{}_{\rm R}$ studied in section~2, the values of $r^{}_2$ versus $r^{}_3$ that lead to successful leptogenesis for several benchmark values of $k$.}
\end{figure*}

\section{$M^{}_{\rm D} \sim {\rm diag}(m^{}_u, m^{}_c, m^{}_t)$}

In this section, we investigate the special case where $M^{}_{\rm D} \sim {\rm diag}(m^{}_u, m^{}_c, m^{}_t)$. This choice is motivated by the fact that in certain SO(10) GUT models, the Dirac neutrino mass matrix approximates the up-type quark mass matrix (see e.g., Refs.~\cite{GUT1}--\cite{GUT4}). This study thus serves to examine the compatibility of the two-zero textures of $M^{}_{\rm R}$ with leptogenesis within the framework of such SO(10) GUT models.

Given that $M^{}_\nu$ is fully determined by the two-zero cofactor conditions, the choice of $M^{}_{\rm D} = {\rm diag}(m^{}_u, m^{}_c, m^{}_t)$ fixes the three right-handed neutrino masses (ordered by increasing mass) to the following values:
\begin{eqnarray}
&& A^{}_1 ({\rm NO}) : \hspace{0.3cm} M^{}_1 \approx M^{}_2 \simeq 3.5 \times 10^{10} \ {\rm GeV} \;, \hspace{0.2cm} M^{}_3 \simeq 6.3 \times 10^{10} \ {\rm GeV} \;; \nonumber \\
&& A^{}_2 ({\rm NO}): \hspace{0.3cm} M^{}_1 \approx M^{}_2 \simeq 2.4 \times 10^{8} \ {\rm GeV} \;, \hspace{0.2cm} M^{}_3 \simeq 1.3 \times 10^{15} \ {\rm GeV}  \;; \nonumber \\
&& B^{}_3 ({\rm NO}): \hspace{0.3cm}  M^{}_1 \simeq 8.0 \times 10^{4} \ {\rm GeV} \;, \hspace{0.2cm} M^{}_2 \approx M^{}_3 \simeq 3.6 \times 10^{12} \ {\rm GeV} \;; \nonumber \\
&& B^{}_3 ({\rm IO}): \hspace{0.3cm}  M^{}_1 \simeq 5.5 \times 10^{4} \ {\rm GeV} \;,  \hspace{0.2cm} M^{}_2 \approx M^{}_3 \simeq 3.3 \times 10^{12} \ {\rm GeV} \;; \nonumber \\
&& B^{}_4 ({\rm NO}): \hspace{0.3cm}  M^{}_1 \simeq 6.2 \times 10^{4} \ {\rm GeV} \;, \hspace{0.2cm} M^{}_2 \simeq 9.0 \times 10^{10} \ {\rm GeV} \;, \hspace{0.2cm} M^{}_3 \simeq 9.3 \times 10^{13} \ {\rm GeV} \;; \nonumber \\
&& B^{}_4 ({\rm IO}): \hspace{0.3cm}  M^{}_1 \simeq 2.5 \times 10^{4} \ {\rm GeV} \;, \hspace{0.2cm} M^{}_2 \simeq 1.8 \times 10^{11} \ {\rm GeV} \;, \hspace{0.2cm} M^{}_3 \simeq 1.0 \times 10^{13} \ {\rm GeV}  \;; \nonumber \\
&& B^{}_6 ({\rm NO}): \hspace{0.3cm} M^{}_1 \simeq 7.5 \times 10^{4} \ {\rm GeV} \;, \hspace{0.2cm} M^{}_2 \simeq 7.6 \times 10^{10} \ {\rm GeV} \;, \hspace{0.2cm} M^{}_3 \simeq 1.5 \times 10^{14} \ {\rm GeV}  \;.
\label{2.2.1}
\end{eqnarray}
For $A^{}_1$ and $A^{}_2$, $N^{}_1$ and $N^{}_2$ are nearly degenerate and lie within the two-flavor regime. In this case, their contributions to the baryon asymmetry should be considered altogether:
\begin{eqnarray}
Y^{}_{\rm B} \simeq  c r \left[ (\varepsilon^{}_{1 e} + \varepsilon^{}_{1 \mu} + \varepsilon^{}_{2 e} + \varepsilon^{}_{2 \mu}) \kappa(K^{}_{1e} + K^{}_{1\mu} + K^{}_{2e} + K^{}_{2\mu}) + (\varepsilon^{}_{1 \tau} + \varepsilon^{}_{2 \tau} ) \kappa(K^{}_{1\tau} + K^{}_{2\tau}) \right] \;,
\label{2.2.2}
\end{eqnarray}
and the expressions for the CP asymmetries are modified accordingly \cite{resonant1, resonant2}
\begin{eqnarray}
\varepsilon^{}_{I\alpha} = \frac{{\rm Im}\left\{ (M^{\prime*}_{\rm D})^{}_{\alpha I} (M^{\prime}_{\rm D})^{}_{\alpha J}
\left[ M^{}_J (M^{\prime\dagger}_{\rm D} M^{\prime}_{\rm D})^{}_{IJ} + M^{}_I (M^{\prime\dagger}_{\rm D} M^{\prime}_{\rm D})^{}_{JI} \right] \right\} }{8\pi  v^2 (M^{\prime\dagger}_{\rm D} M^{\prime}_{\rm D})^{}_{II}} \cdot \frac{M^{}_I \Delta M^2_{IJ}}{(\Delta M^2_{IJ})^2 + M^2_I \Gamma^2_J} \;,
\label{2.2.3}
\end{eqnarray}
where $\Delta M^2_{IJ} \equiv M^2_I - M^2_J$ has been defined and $\Gamma^{}_J= (M^\dagger_{\rm D} M^{}_{\rm D})^{}_{JJ} M^{}_J/(8\pi v^2)$ denotes the decay rate of $N^{}_J$ (for $J \neq I$). Nevertheless, our numerical results indicate that $Y^{}_{\rm B}$ is only about $1.3 \times 10^{-14}$ for  $A^{}_1$ and $6.1 \times 10^{-13}$ for $A^{}_2$, which are respectively smaller than the observed value by approximately four and two orders of magnitude.

For $B^{}_3$, $B^{}_4$ and $B^{}_6$, $N^{}_1$ lies far below the Davidson-Ibarra bound $10^9$ GeV \cite{DI}, implying that leptogenesis driven by $N^{}_1$ alone cannot account for the observed baryon asymmetry. As demonstrated in Ref.~\cite{GUT}, this issue can be resolved by properly incorporating leptogenesis from the next-to-lightest right-handed neutrino $N^{}_2$ and flavor effects \cite{flavor1,flavor2}.
Specifically, for $B^{}_3$ where $N^{}_2$ and $N^{}_3$ are nearly degenerate and lie within the unflavored regime, the resulting baryon asymmetry from their decays can be calculated as
\begin{eqnarray}
Y^{}_{\rm B} \simeq  c r \left[ (\varepsilon^{}_{2 e} +  \varepsilon^{}_{3 e} ) e^{-{3\pi\over 8}\,K^{}_{1 e}} + ( \varepsilon^{}_{2 \mu} +  \varepsilon^{}_{3 \mu})  e^{-{3\pi\over 8}\,K^{}_{1 \mu}} + (\varepsilon^{}_{2 \tau} + \varepsilon^{}_{3 \tau} ) e^{-{3\pi\over 8}\,K^{}_{1 \tau}} \right] \kappa(K^{}_{2} + K^{}_{3})  \;,
\label{2.2.4}
\end{eqnarray}
where the exponential factors account for the washout effects induced by $N^{}_1$.
Nevertheless, our numerical results show that the baryon asymmetry produced by $N^{}_2$ and $N^{}_3$ is smaller than the observed value by several tens of orders of magnitude, owing to the strong washout effects induced by $N^{}_1$.

For $B^{}_4$ and $B^{}_6$ where $N^{}_2$ lies within the two-flavor regime, the baryon asymmetry generated from its decays can be calculated as
\begin{eqnarray}
Y^{}_{\rm B} \simeq cr \left[ \varepsilon^{}_{2 e} \kappa(K^{}_{2e} + K^{}_{2\mu}) e^{-{3\pi\over 8}\,K^{}_{1 e}} + \varepsilon^{}_{2 \mu} \kappa(K^{}_{2e} + K^{}_{2\mu}) e^{-{3\pi\over 8}\,K^{}_{1 \mu}} + \varepsilon^{}_{2 \tau} \kappa(K^{}_{2\tau} ) e^{-{3\pi\over 8}\,K^{}_{1 \tau}} \right] \;,
\label{2.2.5}
\end{eqnarray}
where the washout effects induced by $N^{}_1$ are also included. Nevertheless, our numerical results show that the baryon asymmetry produced by $N^{}_2$ in these scenarios is at most of order $10^{-14}$, far below the observed value.

However, in realistic SO(10) GUT models, $M^{}_{\rm D}$ is not strictly identical to the up-type quark mass matrix but only approximates it. It is therefore more theoretically consistent to parameterize $M^{}_{\rm D}$ as
\begin{eqnarray}
M^{}_{\rm D} = \left( \begin{matrix}
\alpha^{}_1 m^{}_u &  &  \cr
 & \alpha^{}_2 m^{}_c  &  \cr
 &  & \alpha^{}_3 m^{}_t
\end{matrix} \right) \;,
\label{2.2.5}
\end{eqnarray}
following the prescription in Ref.~\cite{GUT}, where the coefficients $\alpha^{}_i$ are ${\mathcal O}(1)$ parameters. Here, we allow these coefficients to vary in the range $0.1 - 10$ and investigate whether the resulting baryon asymmetry can be enhanced to match the observed value.
Our results fall into three distinct categories: (i) For $A^{}_1$, appropriate choices of the $\alpha^{}_i$ coefficients yield $Y^{}_{\rm B}$ that successfully reproduces the observed baryon asymmetry. (ii) For $A^{}_2$ and $B^{}_3$ in the NO case, the maximum achievable $Y^{}_{\rm B}$ is close to the observed value, with $Y_{\rm B}^{\rm max} \simeq 6.6 \times 10^{-11}$ and $3.1 \times 10^{-11}$, respectively. (iii) For all remaining cases, the maximum $Y^{}_{\rm B}$ is suppressed by at least one order of magnitude, and in many instances by many orders of magnitude, falling far short of the observed value.

\section{$M^{}_{\rm D} \propto I$}

Finally, we consider the special case where $M^{}_{\rm D} \propto I$. This choice is motivated by the fact that in certain non-Abelian flavor symmetry models $M^{}_{\rm D}$ is naturally proportional to the identity matrix (see e.g., Ref.~\cite{FS}). This study thus serves to examine the compatibility of the two-zero textures of $M^{}_{\rm R}$ with leptogenesis within the framework of non-Abelian flavor symmetries.

As mentioned above, for the scenario with a non-diagonal $M^{}_{\rm R}$, it is more convenient to perform the leptogenesis calculations in the basis where $M^{}_{\rm R}$ is diagonalized by a unitary transformation $V^{}_{\rm R}$. In this basis, the Dirac neutrino mass matrix is rotated to $M^{\prime}_{\rm D} = M^{}_{\rm D} V^{}_{\rm R}$. For the present case with $M^{}_{\rm D} \propto I$, $M^{\prime}_{\rm D}$ is simply proportional to $V^{}_{\rm R}$. In this scenario, leptogenesis cannot proceed, as the CP asymmetries for the decays of right-handed neutrinos vanish (see Eq.~(\ref{2.8}) for illustration). Refs.~\cite{rge1}-\cite{rge4} have shown that the renormalization group evolution effects can break this relation slightly, thereby enabling successful leptogenesis. We will demonstrate that this issue can also be resolved by embedding the model within a left-right symmetric framework.

\section{Summary}

In this work, we investigate the leptogenesis implications of two-zero textures of the Majorana mass matrix for the right-handed neutrinos $M^{}_{\rm R}$ in the type-I seesaw model, where $M^{}_{\rm R}$ is more fundamental than the light neutrino mass matrix $M^{}_\nu$ and texture zeros on $M^{}_{\rm R}$ can naturally originate from underlying flavor symmetries. We take the Dirac neutrino mass matrix $M^{}_{\rm D}$ as diagonal, and carry out the study in three scenarios: a general diagonal $M^{}_{\rm D}$, the SO(10) GUT-inspired $M^{}_{\rm D} \sim \text{diag}(m^{}_u, m^{}_c, m^{}_t)$ and the non-Abelian flavor symmetry-motivated $M^{}_{\rm D} \propto I$.

For the general diagonal $M^{}_{\rm D} = k\text{diag}(1, r_2, r_3)$, we first derive the two-zero cofactor conditions of $M_\nu$ (equivalent to two-zero textures of $M^{}_{\rm R}$) and select five experimentally viable $M^{}_{\rm R}$ texture classes ($A^{}_1$, $A^{}_2$, $B^{}_3$, $B^{}_4$ and $B^{}_6$) consistent with the latest neutrino oscillation data within $3\sigma$ ranges. We give predictions for the undetermined neutrino parameters (the lightest neutrino mass and three CP phases) for these classes, and calculate the lightest right-handed neutrino mass $M^{}_1$, finding that $M^{}_1$ scales with $k^2$ and increases with $r^{}_2$ and $r^{}_3$. We then employ flavor-dependent leptogenesis formulas appropriate for different temperature regimes to carry out leptogenesis analysis, and determine the regions of the parameters $k$, $r^{}_2$ and $r^{}_3$ that allow for successful leptogenesis.

For the SO(10)-inspired $M^{}_{\rm D} \sim \text{diag}(m^{}_u, m^{}_c, m^{}_t)$, we fix the masses of three right-handed neutrinos for all viable texture classes.
We find that $N^{}_1$ and $N^{}_2$ are nearly degenerate for $A^{}_1$ and $A^{}_2$, yet they still fail to generate a sufficient baryon asymmetry; for $B^{}_3$, $B^{}_4$ and $B^{}_6$, $N^{}_1$ lies far below the Davidson-Ibarra bound and is thus incapable of producing the observed baryon asymmetry.
We further find that this issue cannot be resolved by accounting for the joint contributions of the heavier right-handed neutrinos ($N^{}_2$/$N^{}_3$) and flavor effects.
In contrast, for the generalized scenario with $M^{}_{\rm D}= \text{diag}(\alpha^{}_1 m^{}_u, \alpha^{}_2 m^{}_c, \alpha^{}_3 m^{}_t)$, which is relevant for realistic SO(10) GUT models, class $A^{}_1$ can achieve successful leptogenesis with appropriate choices of the $\mathcal{O}(1)$ parameters $\alpha^{}_i$, while the remaining cases still fail to yield the observed baryon asymmetry. In future work, we will embed the present framework into SO(10) GUT models equipped with the left-right symmetric structure.

For the non-Abelian flavor symmetry-inspired $M^{}_{\rm D} \propto I$, the rotated Dirac mass matrix $M^{\prime}_{\rm D} = M^{}_{\rm D} V^{}_{\rm R}$ is proportional to the unitary matrix $ V^{}_{\rm R}$ that diagonalizes $M^{}_{\rm R}$, leading to vanishing CP asymmetries for right-handed neutrino decays and thus invalid leptogenesis. We point out that embedding the model into the left-right symmetric framework is a feasible solution, which will be verified with detailed calculations in future work.

Our study establishes the close connection between two-zero textures of the fundamental $M^{}_{\rm R}$ and leptogenesis in the type-I seesaw model, and the distinct phenomenological predictions of different texture classes can be tested by future neutrino oscillation experiments. This work provides important implications for exploring the neutrino mass generation mechanism and the underlying flavor theory beyond the SM.

\vspace{0.5cm}

\underline{Acknowledgments} \vspace{0.2cm}

This work was supported in part by the National Natural Science Foundation of China under Grant
No.~12475112, Liaoning Revitalization Talents Program under Grant No.~XLYC2403152, and the
Basic Research Business Fees for Universities in Liaoning Province under Grant No.~LJ212410165050.

\end{document}